\documentstyle[multicol,aps,prb,epsf]{revtex}
\newcommand{\Vec}[1]{\mbox{\boldmath$#1$}}
\begin{document}

\draft

\title{
High temperature superconductivity in dimer array systems
}

\author{
Kazuhiko Kuroki$^1$, Takashi Kimura$^2$, and Ryotaro Arita$^3$ 
}

\address{
$^1$ Department of Applied Physics and Chemistry, the University of
Electro-communications, Chofugaoka, Chofu-shi, Tokyo 182-8585, Japan\\
$^2$ Advanced Research Institute for Science and Engineering, 
Waseda University, Ohkubo, Shinjuku-ku, Tokyo 169-8555, Japan\\
$^3$ Department of Physics, University of Tokyo, Hongo, Tokyo 113-0033, Japan\\
}

\date{\today}

\maketitle

\begin{abstract}
Superconductivity in 
the Hubbard model is studied on a series of lattices in which dimers are 
coupled in various types of arrays.
Using fluctuation exchange method and solving the linearized 
Eliashberg equation, the transition temperature $T_c$ of these 
systems is estimated to be much higher than that of the Hubbard model on
a simple square lattice, which is a model for the high $T_c$ cuprates. 
We conclude that these `dimer array' systems
can generally exhibit superconductivity with very high $T_c$.
Not only $d$-electron systems, but 
also $p$-electron systems may provide various stages for 
realizing the present mechanism.
\end{abstract}

\medskip


\begin{multicols}{2}
\narrowtext
Search for high temperature superconductivity is one of the 
most challenging problems in solid state physics.
Since one of the bottleneck for high $T_c$ is the generally low 
energy scale of the phonons, purely electronic pairing mechanism 
have attracted much attention from the early days.
The discovery of high $T_c$ superconductivity in the cuprates\cite{BM}
has certainly boosted interest along this line.
In particular, the possibility of superconductivity in 
the Hubbard model on a square lattice, a simplest purely 
electronic model for the cuprates, 
has been intensively studied, where spin fluctuation theories using
fluctuation exchange (FLEX) method 
have estimated $T_c$ of the $d_{x^2-y^2}$-wave superconductivity
to be $O(0.01t)$ ($t$ is the nearest neighbor hopping integral).
\cite{Bickers,Dahm,Grabowski} 
Since $t\sim 0.4$eV for the cuprates, 
this estimation is consistent with the experimentally observed 
$T_c$ of up to $150$K, 
but at the same time one should note that it is {\it two orders of 
magnitude lower} than the kinetic energy scale $t$.

In our view, 
one reason for this reduction of $T_c$ is that the $d_{x^2-y^2}$ 
gap function changes sign, having nodal lines that intersect the Fermi 
surface. Such a sign change in the gap function is 
a `necessary evil' for spin fluctuation pairing 
in that the pair scattering processes 
$[\Vec{k}\uparrow,\Vec{-k}\downarrow]\rightarrow [\Vec{k+Q}\uparrow,\Vec{-k-Q}\downarrow]$,
mediated by the dominant spin fluctuations with wave vector $\Vec{Q}$,
have to accompany a sign change in the gap function $\phi$, i.e., 
$\phi(\Vec{k})\phi(\Vec{k+Q})<0$.
In the case of the square lattice, this requirement 
leads to the $d_{x^2-y^2}$ pairing, where the gap function has nodes 
that intersect the Fermi surface, thereby resulting in a reduction of $T_c$. 

\begin{figure}
\begin{center}
\leavevmode\epsfysize=30mm \epsfbox{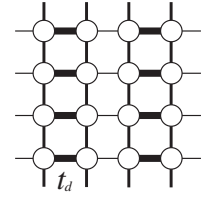}
\caption{The two band lattice studied in  ref.\protect\onlinecite{KAr}.
The thickness of the lines represent the magnitude of hopping integrals.}
\label{fig1}
\end{center}
\end{figure}

From this viewpoint, two of the present authors have recently proposed that
superconductivity with much higher $T_c$ can be achieved in 
systems having two pocket-like nested Fermi surfaces, where 
the dominant pair scattering processes take place {\it between}
the two Fermi surfaces, so that 
the sign change in the gap function takes place {\it across}, not {\it on}, 
the Fermi surfaces.\cite{KAr}
Along this line, Hubbard models on a two band lattice 
(Fig.\ref{fig1})\cite{KAr} and a 
\noindent four band one\cite{Takashi} have been studied using FLEX method, 
where high $T_c$ has been obtained solving the linearized Eliashberg
equation. 

If we focus on the two band model shown in Fig.\ref{fig1}, 
the obtained gap function indeed changes sign across the two bands 
but stays nearly constant within each band.\cite{KAr}
Such a form of the gap function can be interpreted 
in terms of a real space picture 
by looking at the system as {\it dimers coupled in an array}.
Namely, it is natural to consider that the singlet pairs are mainly 
formed between electrons residing on different sites within a dimer 
due to the antiferromagnetic interaction $\propto t_d^2/U$ 
between the two sites.
Then, the gap function has $s$ wave symmetry within each band because the 
pairing occurs within a unit cell,
while it changes sign across the two bands because the pairs 
are formed between different sites.\cite{commentgap}

This intuitive picture has led us to consider here the Hubbard model on 
a series of lattices in which dimers are coupled in various types of arrays.
We conclude that these `dimer array' systems can generally  
exhibit superconductivity with very high $T_c$.

Let us now start with a dimer array Hubbard system 
\begin{eqnarray*}
H&=&
-t_d\sum_{i,\sigma}(c^\dagger_{i\sigma}d_{i\sigma}+
d^\dagger_{i\sigma}c_{i\sigma})
-t'\sum_{\langle i,j\rangle,\sigma}
(c^\dagger_{i\sigma}c_{j\sigma}\\&&+d^\dagger_{i\sigma}d_{j\sigma}
+{\rm H.c})
+U\sum_i(n^c_{i\uparrow}n^c_{i\downarrow}+
n^d_{i\uparrow}n^d_{i\downarrow})
\end{eqnarray*}
\begin{figure}
\begin{center}
\leavevmode\epsfysize=50mm \epsfbox{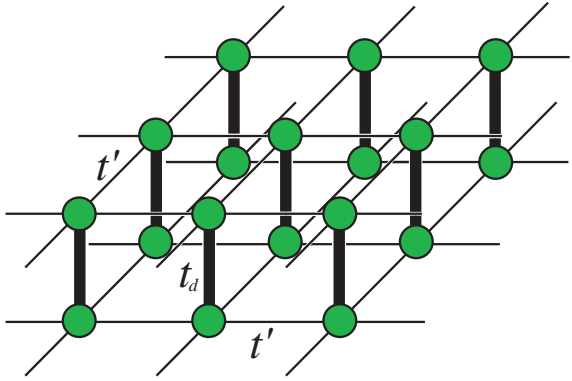}
\caption{Dimer array on a square lattice.}
\label{fig2}
\end{center}
\end{figure}
\noindent
on a square lattice shown in Fig.\ref{fig2}.\cite{commDagotto} 
Here, $c_{i\sigma}^\dagger/c_{i\sigma}/n^c_{i\sigma}$ 
and $d_{i\sigma}^\dagger/d_{i\sigma}/n^d_{i\sigma}$ 
denote the creation/annhilation/number operators on the lower and upper sites 
of the $i$-the dimer, respectively. We assume $t_d>t'$, 
where $t_d$ is the hopping within the 
dimers, while the hopping $t'$ couples the dimers. 
When $U=0$, the bonding and antibonding bands have band width of 
$8t'$ each, while the level offset between the bands is $2t_d$. 
So as far as $t_d$ is not too large compared to $t'$, 
the two bands have a certain amount of overlap. 
Then, if the band filling $n$\cite{commentn} is not 
too far away from half-filling ($n=1$), both bands cross the Fermi 
level to result in two pocket-like Fermi surfaces, 
one of which lies around $\Vec{k}=(0,0)$ and the other around 
$\Vec{k}=(\pi,\pi)$ (Fig.\ref{fig3}).
The two Fermi surfaces are nested to some degree 
when $n$ is close to half-filling, so that when $U$ is turned on, 
antiferromagnetic spin fluctuations 
that mediate inter-Fermi-surface pair scattering 
(arrows in (Fig.\ref{fig3})) arises,
which should lead to a superconductivity with a 
gap function that changes sign across the Fermi surfaces  
but not within each Fermi surface.

On the other hand, $t_d$ should not be too small, namely,
\begin{figure}
\begin{center}
\leavevmode\epsfysize=60mm \epsfbox{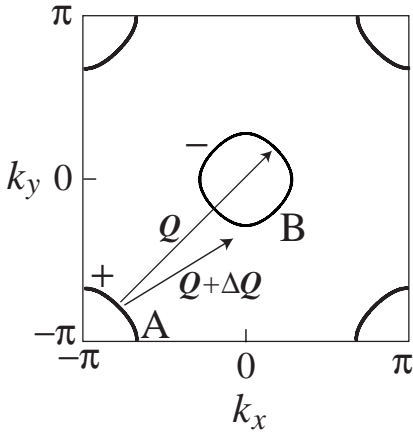}
\caption{Fermi surfaces of the dimer array system on a square lattice.
$\Vec{Q}$ denotes the nesting vector, i.e., the peak position of the 
spin susceptibility, while 
$\Delta\Vec{Q}$ represents the spread of the spin susceptibility around 
$\Vec{Q}$.}
\label{fig3}
\end{center}
\end{figure}

\noindent the overlap of the two bands, and 
thus the size of the Fermi surfaces, should not be too large 
for high $T_c$ superconductivity to occur.
This is because when the Fermi surfaces are small compared to 
the spread $\Delta Q$ of the peak in the spin susceptibility,
inter-Fermi-surface pair scatterings 
$\forall[\Vec{k}\uparrow, \Vec{-k}\downarrow]\in {\rm A}\rightarrow\forall[\Vec{k+q}\uparrow, \Vec{-k-q}\downarrow]\in {\rm B}$
all have large contribution,
while when the Fermi surfaces are large compared to $\Delta Q$, the 
pair scattering processes having large contribution will be 
restricted to a certain combination of initial states $\Vec{k}\in {\rm A}$ and 
final states $\Vec{k+q}\in {\rm B}$.\cite{KAr}

In order to verify the above expectation, 
we use the multiband version\cite{Takashi,KU} of FLEX\cite{Bickers},
which is kind of a self-consistent random phase approximation,
to obtain the renormalized Green's function of the Hubbard model.
Then, $T_c$ is determined as the temperature at 
which the eigenvalue $\lambda$ of the Eliashberg equation, 
\begin{eqnarray*}
\lambda\phi_{l m}(k)\nonumber &=& -\frac{T}{N}\sum_{k'}\sum_{l',m'}\\
&&\times V_{l m}(k-k')G_{ll'}(k')G_{mm'}(-k')\phi_{l'm'}(k'),
\end{eqnarray*}
reaches unity.
Here, $G$ is the renormalized Green's function matrix 
(with $l,m,$etc... labeling sites in a unit cell) obtained by FLEX.
$V$ is the spin singlet pairing interaction matrix 
given by $V(q)=\frac{3}{2}V_{\rm sp}(q)-\frac{1}{2}V_{\rm ch}(q)$, 
where the pairing interaction due to spin fluctuations (sp) and that 
due to charge fluctuations (ch) are given as 
$V_{\rm sp(ch)}=U^2\chi_{\rm sp(ch)}$
using the spin and charge susceptibilities 
$\chi_{\rm sp(ch)}(q)=\chi_{\rm irr}(q)[1-(+)U\chi_{\rm irr}(q)]^{-1}$,
where $\chi_{\rm irr}(q)$ is the irreducible susceptibility matrix 
$\chi_{\rm irr}(q)= -\frac{1}{N}\sum_k G(k+q)G(k)$ 
($N$ is the number of $k$-point meshes).
In the present paper, we take up to $64\times 64$ $k$-point meshes and 
up to 4096 Matsubara frequencies. 

As mentioned above, the FLEX+Eliashberg equation approach 
is known to give reasonable $T_c$ of order $0.01t$ 
for the single band Hubbard model\cite{Bickers,Dahm,Grabowski}, 
which is a model for the high $T_c$ cuprates. The occurrence of 
superconductivity in the Hubbard model contradicts with some of 
the numerical studies, but we believe this to be due to finite 
size effects in the numerical calculations. In fact, 
a quantum Monte Carlo calculation that 
pays special attention to the discreteness of the level spacing 
in finite size systems has detected large enhancement of the pairing 
correlation function, which is at least qualitatively consistent 
with the FLEX results.\cite{KAo}

\begin{figure}
\begin{center}
\leavevmode\epsfysize=50mm \epsfbox{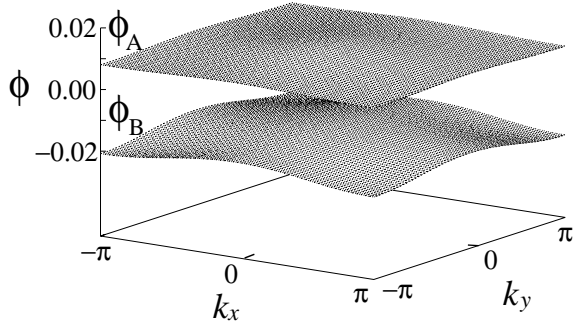}
\caption{The gap functions for the two bands A and B with 
$U=8$, $n=0.95$, $t'=0.8$, $t_d=1.3$, and $T=0.15$.}
\label{fig4}
\end{center}
\end{figure}

\begin{figure}
\begin{center}
\leavevmode\epsfysize=80mm \epsfbox{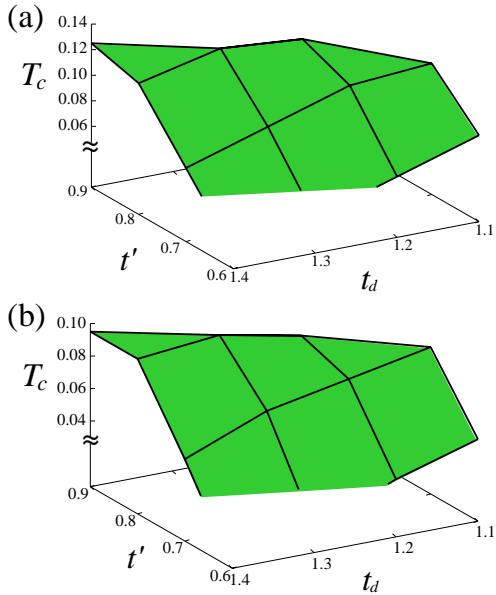}
\caption{$T_c$ of the dimer array Hubbard model on a square lattice 
as a function of $(t',t_d)$ with $U=8$, $n=0.95$(a), or $n=0.9$(b). }
\label{fig5}
\end{center}
\end{figure}

We have performed FLEX calculation for $n=0.95$ and $n=0.9$ with $U=8$, 
varying $(t',t_d)$ in a certain range. 
The gap function obtained by solving the linearized Eliashberg equation 
(Fig.\ref{fig4}) does not change sign in each band, while it 
changes sign across the two bands, as expected.
As a consequence, $T_c$, shown in Fig.\ref{fig5}, 
reaches $\sim 0.13$ for $n=0.95$, 
which is $4\sim 5$ times higher than the typical $T_c$ 
of the Hubbard model, estimated by the same way, 
on a simple square lattice.\cite{Bickers,Dahm,Grabowski} 

As $t'/t_d$ increases, the Fermi surfaces
become large, resulting in a reduction of $\lambda$ due to the reason 
mentioned above. 
At the same time, the antiferromagnetic spin fluctuations increases 
rapidly because 
the nesting between the two Fermi surfaces becomes better for large Fermi 
surfaces.
Thus, the tendency towards magnetic ordering
dominates over superconductivity. On the other hand, when $t'/t_d$ is 
small, only one band intersects the Fermi level, so that the Fermi surface
nesting is degraded and $T_c$ becomes low. Consequently, high $T_c$
is obtained in a certain optimized regime of $(t',t_d)$ as seen in 
Fig.\ref{fig5}. At $n=0.9$, $T_c$ becomes slightly lower, 
but still reaches $0.1$.
\begin{figure}
\begin{center}
\leavevmode\epsfysize=85mm \epsfbox{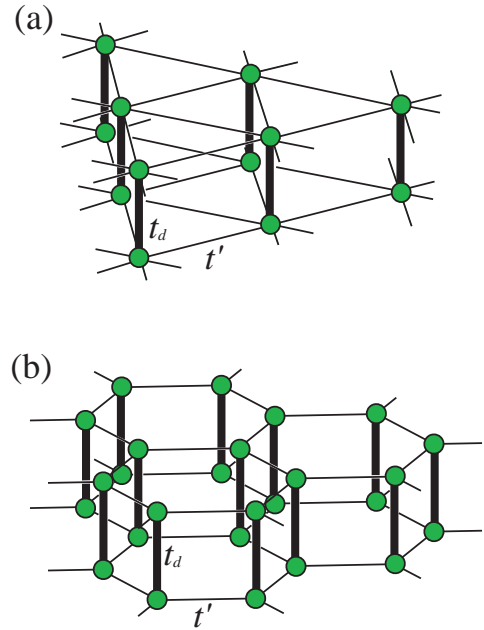}
\caption{Dimer array on triangular (a) and honeycomb (b) lattices.}
\label{fig6}
\end{center}
\end{figure}

\begin{figure}
\begin{center}
\leavevmode\epsfysize=85mm \epsfbox{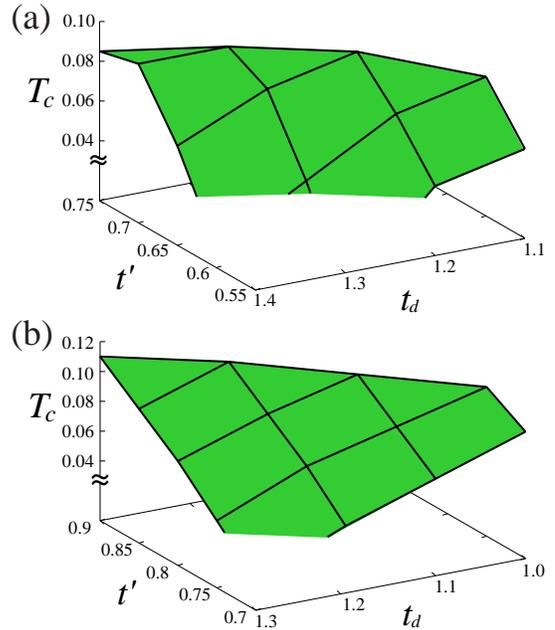}
\caption{$T_c$ of the dimer array systems on 
triangular (a) or honeycomb (b) lattices with $U=8$ and $n=0.95$.}
\label{fig7}
\end{center}
\end{figure}

We now show that high $T_c$ superconductivity in dimer array systems is quite 
general by looking into triangular and honeycomb lattices
shown in Fig.\ref{fig6}(a) and (b), respectively.
$T_c$ for $n=0.95$ and $U=8$ on these lattices 
are shown in Fig.\ref{fig7} as functions of $(t',t_d)$, 
which again exceed or come close to $0.1$. 
 These results suggest that 
dimer array systems can generally exhibit high $T_c$ 
superconductivity on various types of lattices.
\begin{figure}
\begin{center}
\leavevmode\epsfysize=70mm \epsfbox{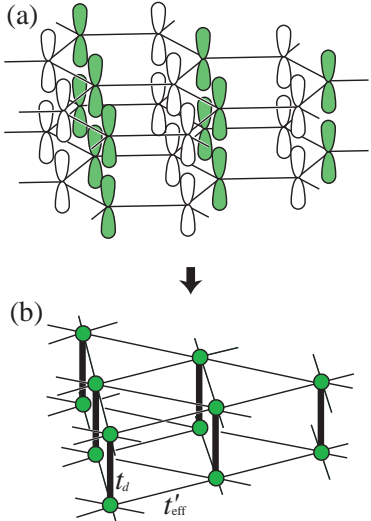}
\caption{(a)Dimer array system (consisting of $p_\pi$ orbitals here) on a 
honeycomb lattice in which the site energy of the unhatched orbitals 
is higher by $\varepsilon$ than that of the hatched ones.
(b)The effective lattice of (a).}
\label{fig8}
\end{center}
\end{figure}

\begin{figure}
\begin{center}
\leavevmode\epsfysize=50mm \epsfbox{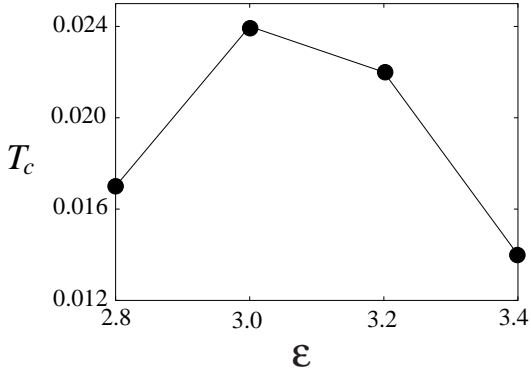}
\caption{$T_c$ of the system shown in Fig.\ref{fig8}(a) 
as a function of $\varepsilon$ with $U=2$, $t'=1$, $t_d=0.5$ and $n=0.5$.}
\label{fig9}
\end{center}
\end{figure}

So far we have focused on parameter regimes with large 
$U/t'$ and $U/t_d$, where we have $d$-electron\cite{commentdz2} 
systems in mind. Namely, if $t'$ and $t_d$ are few hundred meV, 
which is typical for $d$-electron systems, 
$T_c\sim 0.1t'$ corresponds to a very high temperature of 
few hundred K.
Then our next question is : can we obtain similarly high $T_c$ in systems with 
smaller $U/t'$, as in $p$-electron systems ?
In $2p$-electron systems such as B, C, and/or N compounds, 
the on site Coulomb repulsion 
and the hopping integrals are both typically  few eV,
so that $U/t\sim O(1)$. 

In such a case, we have found that $T_c$ estimated by 
FLEX+Eliashberg equation is very low (or does not exist) 
not only on conventional lattices,  
but also on the lattices considered above as well.
Nevertheless, we now show that high $T_c$ can be achieved even for 
systems with small $U/t'$ by introducing a level offset between 
neighboring sites, thus reducing the effective band width.
As an example, let us consider a dimer array system on a 
honeycomb lattice where the site energy differs by $\varepsilon$
between neighboring sites (Fig.\ref{fig8}(a)).\cite{commentBN} 
At quarter filling ($n=0.5$) and 
in the limit of large $\varepsilon$, the system becomes equivalent to
a half-filled dimer array system on a triangular lattice with 
an effective hopping $t'_{\rm eff}=t^2/\varepsilon$ between 
the dimers (Fig.\ref{fig8}(b)).
Consequently, we can expect high $T_c$ when the value of $\varepsilon$
is tuned so that $U/t'_{\rm eff}$ is optimized.
In Fig.\ref{fig9}, we show $T_c$ for $U=2$, $t'=1$, $t_d=0.5$
\cite{commentMgB2} and 
$n=0.5$ as a function of $\varepsilon$. $T_c$ is found to exceed 0.02, which 
again corresponds to few hundred K  
if $t'=1$ corresponds to few eV as in $2p$ systems.

To summarize, we have shown that `dimer array' Hubbard systems can generally 
exhibit superconductivity with very high $T_c$. Not only $d$-electron
systems, but also $p$-electron systems may provide various stages 
for realizing the present mechanism.

We would like to acknowledge H. Aoki, J. Akimitsu, and Y. Zenitani for
valuable discussions. Numerical calculations where performed at the
Supercomputer Center, Institute for Solid Stated Physics, University of 
Tokyo.

\end{multicols}

\begin{references}
\bibitem{commentgap} The creation operators of 
bonding (A) and antibonding (B) orbitals
can be given as $c^\dagger_{\rm A(B)}
\propto c^\dagger_{\rm 1}+(-)c^\dagger_{\rm 2}$ in terms of the 
creation operators on sites 1 and 2 within a dimer.
If the pairing occurs only between sites 1 and 2
(i.e., $\langle c^\dagger_1c^\dagger_1\rangle
=\langle c^\dagger_2c^\dagger_2\rangle=0$), 
then $\langle c_{\rm A}^\dagger  c_{\rm A}^\dagger\rangle=-
\langle c_{\rm B}^\dagger  c_{\rm B}^\dagger\rangle$ holds.
\bibitem{commentn} Throughout the paper, the band filling is defined as 
$n=$[number of electrons]/[number of sites].
\bibitem{commentdz2} $d_{3z^2-r^2}$ orbital may be suitable for our purpose 
since we need large hopping in the out-of-plane direction.
\bibitem{commentBN} Such an alternation of the $2p$ orbital energy in 
a honeycomb plane is realized, for example, in BN graphite.
\bibitem{commentMgB2} In taking $t_d/t'=0.5$,
we have referred to the ratio between inter- and intraplane hoppings of 
B$2p_\pi$ orbitals in MgB$_2$.
\end{references}
\end{document}